\documentclass[doublecol]{epl2} 
\usepackage{amsfonts}
\usepackage{amssymb}
\usepackage{amsmath, mathtools,latexsym}
\usepackage{graphicx}
\usepackage{color}
\usepackage{enumerate}

\newcommand{\ket}[1]{|#1\rangle}
\newcommand{\bra}[1]{\langle#1|}

\newcommand{\supN}{^{(N)}}  
\newcommand{\mathbox}[4][\normalsize]{\makebox[#2][#3]{#1$#4$}}

\usepackage{booktabs,array}
\makeatletter
\renewcommand{\thepage}{\arabic{page}}
\renewcommand{\ps@eplfirst}{\renewcommand{\@oddhead}{}}
\makeatother
\title{Implementing a neutral-atom controlled-phase gate with a single Rydberg pulse}

\author{Rui Han\inst{1,2,3} \and Hui Khoon Ng\inst{1,4,5} \and Berthold-Georg Englert\inst{1,5,6}}
\shortauthor{R. Han \etal}

\institute{                    
  \inst{1} Centre for Quantum Technologies, National University of
  Singapore - 117543 Singapore\\
  \inst{2} Max Planck Institute for the Science of Light - 91058 Erlangen, Germany, EU\\
  \inst{3} Institute of Optics, Information and Photonics, University of Erlangen-N\"urnberg - 91058 Erlangen, Germany, EU\\
  \inst{4} Yale-NUS College - 138614 Singapore\\
  \inst{5} MajuLab, CNRS-UNS-NUS-NTU International Joint Research Unit, UMI 3654 - Singapore\\
  \inst{6} Department of Physics, National University of
  Singapore - 117542 Singapore
}
\pacs{03.67.Lx}{Quantum computation architectures and implementations}
\pacs{32.80.Qk}{Coherent control of atomic interactions with photons}
\pacs{32.80.Ee}{Rydberg states}

\abstract{
One can implement fast two-qubit entangling gates by exploiting the Rydberg blockade. Although various theoretical schemes have been proposed, experimenters have not yet been able to demonstrate  two-atom gates of high fidelity due to experimental constraints. We propose a novel scheme, which only uses a single Rydberg pulse illuminating both atoms, for the construction of neutral-atom controlled-phase gates. In contrast to the existing schemes, our approach is simpler to implement and requires neither individual addressing of atoms nor adiabatic procedures. With parameters estimated based on actual experimental scenarios, a gate fidelity higher than 0.99 is achievable.}

\begin{document}

\maketitle

\section{Introduction}

Fueled by advances in the experimental techniques of trapping, cooling, and manipulating neutral atoms, neutral-atom qubits are regarded as one of the most promising approaches to quantum computing~\cite{Bloch08, Schlosser11}. One of the key challenges in this approach is the realization of two-qubit entangling gates. In 2000, Jaksch \textit{et al.}~\cite{Jaksch00} proposed schemes for fast neutral-atom quantum gates via Rydberg blockade~\cite{Gallagher94}. Since then, much effort has been put into studies of implementing controlled-NOT (CX) and controlled-phase (Cphase) gates with the help of Rydberg blockade~\textcolor{black}{\cite{Lukin01, Brion07, Urban09, Isenhower10, Zhang10, Saffman10, Isenhower11, Saffman11, Zhang12, Keating13, Xia13, Muller14, Keating15}. Rydberg blockade and collective Rydberg excitation were observed experimentally~\cite{Heidemann07, Urban09, Gaetan09, Barredo14}; in addition, Rydberg gates and entanglement of neutral-atom qubits were also demonstrated~\cite{Wilk10, Zhang10, PhysRevA.92.022336, Jau16}. However, for two neutral-atom qubits, the highest measured entangling gate fidelity or the fidelity of Bell state preparation is about 0.8 after correcting for atom loss}, a huge shortfall from the theoretically expected error of $10^{-3}$~\cite{Saffman05, Keating15}. A practical scheme for a two-atom entangling gate of high fidelity is still lacking.

In theory, with strong Rydberg blockade, the fidelity of these Rydberg two-atom gates could be better than 0.99; in practice, existing proposals suffer from experimental constraints. For schemes that require individual addressing of the atoms, the gate is constructed by applying a sequence of Rydberg pulses, \textit{e.g.}, at least three pulses are needed for a controlled-$Z$ (CZ) gate. Other than the technical challenge of individual addressability --- \textcolor{black}{tackled by placing the atoms at least a few $\mu$m from each other which weakens the blockade effect as an unavoidable consequence} --- the main error arises from the dephasing \textcolor{black}{and population loss} of the Rydberg state during the time gaps between the pulses addressing individual atoms~\cite{Zhang10, Saffman11}.
For symmetric entangling gates (such as Cphase gates), one expects to be able to implement them without distinguishing between the two atoms or individually addressing them.
Indeed, various schemes that do not require individual addressing have been proposed, but they also suffer from a variety of problems. A CZ gate can be constructed by exciting the atoms coherently to the doubly excited Rydberg state~\cite{Jaksch00}, but the resulting large mechanical force is difficult to counter. \textcolor{black}{In the proposals of Refs.~\cite{Brion07} and~\cite{PhysRevA.93.012306} for the CZ gate, the Rydberg states are not substantially populated; these schemes are challenging, however, as they rely on exact knowledge of the Rydberg blockade energy or on efficient high-order multi-photon ground-state to ground-state transitions via the doubly excited Rydberg state.} The adiabatic gate scheme of Ref.~\cite{Jaksch00} was also investigated~\cite{Moller08, Muller14, Keating15}; with numerically optimized yet experimentally feasible parameters, the CZ gate fidelity is limited to 0.983 due to population losses during the adiabatic processes (even when assuming perfect control at zero temperature). To date, no Cphase gate scheme without individual addressing of the atoms has been experimentally demonstrated.

In this paper, we describe a novel scheme that is simpler and more robust than earlier proposals. It implements a
two-atom Cphase gate with a single Rydberg pulse driving both atoms simultaneously and symmetrically. This scheme neither relies on populating the doubly excited Rydberg state nor requires very strong atom-light couplings. 

In the following, we first explain the principle behind our proposal for a CZ gate. Then, the scheme is illustrated by examples with a list of the required parameter values and the minimal achievable gate fidelities. One example, which offers a minimum gate fidelity higher than 0.99, is described in detail. Finally, we demonstrate how to generalize this scheme to an arbitrary Cphase gate. The robustness of this implementation is discussed as well.

\section{The physical system}

\begin{figure}[t]
\centerline{\setlength{\unitlength}{1pt}
\centering
\begin{picture}(220,125)(0,0)
\put(-5,-5){\includegraphics[width=7.5cm]{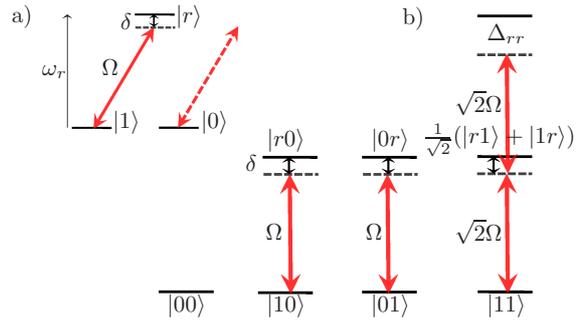}}
\end{picture}}
\caption{The level-diagrams of the interaction between the light field with a) a single atom and b) two atoms. 
Note: This level-diagram is similar to that of Model A in Ref.~\cite{Jaksch00}.}\label{fig1}
\end{figure}

Two degenerate ground states of an atom, labelled by $\ket{0}$ and $\ket{1}$, form the basis of a qubit. We suppose that we can employ suitable levels and light fields such that $\ket{1}$ is coupled to Rydberg state $\ket{r}$ with a coupling strength $\Omega$, while $\ket{0}$ is not coupled to any state
; see fig.~\ref{fig1}a. Under these circumstances, each atom can be treated as a three-level system with levels $\ket{0},\ket{1}$ and $\ket{r}$. Denoting the effective coupling strength between $\ket{1}$ and $\ket{r}$ by $\Omega$ and the detuning by $\delta$, the interaction Hamiltonian for this single-atom system is
\begin{eqnarray}\label{eq1}
H_I^{(1)}\widehat=\frac{\hbar}{2}\left(\begin{array}{cc} -\delta & \Omega  \\\Omega^* & \delta \end{array} \right),
\end{eqnarray}
written in the basis $\{\ket{1},\ket{r}\}$ (as $\ket{0}$ is decoupled), under the rotating-wave approximation. Depending on the choice of Rydberg state and the available experimental set-ups, Rabi oscillations between the ground and Rydberg states can be obtained by either a high-frequency laser beam or a two-photon Raman transition with two laser beams~\cite{Johnson08}. For simplicity, we omit this detail and just consider the overall coupling strength between the ground and Rydberg states denoted by $\Omega$. 

If two atoms in state $\ket{11}$ are excited to the Rydberg state simultaneously and symmetrically, the relevant states are $\{\ket{11},\; (\ket{r1}+\ket{1r})/\sqrt{2},\;\ket{rr}\}$; see fig.~\ref{fig1}b. The interaction Hamiltonian in this basis is
\begin{eqnarray}\label{eq2}
H_I^{(2)}\widehat=\frac{\hbar}{2}\left(\begin{array}{ccc} -\delta & \sqrt{2}\Omega & 0 \\ \sqrt{2}\Omega^* & \delta & \sqrt{2}\Omega \\ 0 & \sqrt{2}\Omega^* & 2\Delta_{rr}+3\delta \end{array} \right),
\end{eqnarray}
where $\Delta_{rr}$ is the energy shift of the doubly excited Rydberg state. In the strong Rydberg-blockade limit, when $|\Delta_{rr}|\gg|\Omega|$, the doubly excited Rydberg state is so far-detuned that it is hardly populated. Thus, one can use adiabatic elimination on the third state and neglect terms of the order $|\Omega/\Delta_{rr}|$ or higher, and obtain a $2\times2$ effective description.
The final state, after a Rydberg excitation pulse of duration $t$, is~\cite{Han13}
\begin{equation}
\ket{\psi(t)}\supN
\widehat{=}{\left(\begin{array}{@{}c@{}} 
\cos{\left(\frac{1}{2}\Omega\supN t\right)}
+i\frac{\delta}{\Omega\supN}\sin{\left(\frac{1}{2}\Omega\supN t\right)} 
\\[0.5ex] 
-i\frac{\Omega^*\sqrt{N}}{\Omega\supN}\sin{\left(\frac{1}{2}\Omega\supN t\right)}
\end{array}\right)},
\end{equation}
where $N=1,2$ is the number of atoms involved in the transition. The column entries are the probability amplitudes for the ground state and the singly excited Rydberg state, respectively. The single-atom and two-atom effective Rabi frequencies for the Rydberg excitation are $\Omega^{(1)}=\sqrt{|\Omega|^2+\delta^2}$ and $\Omega^{(2)}=\sqrt{2|\Omega|^2+\delta^2}$. The evolution for a collective single-Rydberg excitation of two atoms is the same as the Rydberg excitation of one atom, except for the $\sqrt{2}$ enhancement of the atom-light coupling strength. 
The ratio between the two effective Rabi frequencies is in the range $1\leq\Omega^{(2)}/\Omega^{(1)}\leq\sqrt{2}$ for all values of $\Omega$ and $\delta$. 

The two-atom system is governed by the full evolution operator $e^{-iHt/\hbar}$ in the nine-dimensional Hilbert space. The resulting gate operation in the subspace spanned by the four ground states, $\ket{00},\ket{10},\ket{01}$, and $\ket{11}$, is
\begin{equation}
G(t)\widehat{=}{\left(\begin{array}{@{}cccc@{}} 
e^{i\frac{\delta}{2}t}&0&0&0\\
0&\mathbox{5em}{c}{\bra{1}e^{-\frac{i}{\hbar} H_I^{(1)} t}\ket{1}}&0&0\\
0&0& \mathbox{2.5em}{c}{\bra{1}e^{-\frac{i}{\hbar} H_I^{(1)} t}\ket{1}}&0\\
0&0&0&\mathbox{7em}{c}{\bra{11}e^{-\frac{i}{\hbar} H_I^{(2)} t}\ket{11}}
\end{array}\right)}.
\end{equation}
A unitary $G(t)$ would require exact phase factors for the diagonal entries but, besides the first one, the amplitudes of the other three diagonal entries oscillate with their respective Rabi-oscillation frequencies. $G(t)$ can, at best, be a good approximation of a unitary operator for some particular time $t$. Note that the full evolution operator has small non-zero off-diagonal elements not belonging to this $4\times4$ subspace; the effect of those elements will be addressed below.

\section{Constructing a CZ gate} 
For $G(t)$ to be a good Cphase gate, the primary requirement is that the absolute values of its diagonal entries are close to unity. This is easy to fulfill in the far-off-resonant coupling regime, where ${|\delta|\!\gg\!|\Omega|}$ and ${\Omega^{(2)}/\Omega^{(1)}\!\rightarrow\!1}$. In this regime, the population mostly remains in the ground states, and the states $\ket{11}$ and $\ket{10}$ (or $\ket{01}$) pick up phase factors $\exp[i\,\mathrm{sgn}(\delta)(\Omega^{(2)}-|\delta|)t/2]$ and  $\exp[i\,\mathrm{sgn}(\delta)(\Omega^{(1)}-|\delta|)t/2]$ relative to $\ket{00}$. However, the ratio of the two phases is close to two for $|\delta|\!\gg\!|\Omega|$, so that the phase accumulated by $\ket{11}$ is twice that of $\ket{01}$ (or $\ket{10}$). Therefore, the CZ gate cannot be realized in the regime where $|\delta|\!\gg\!|\Omega|$.

\begin{figure}[t]
\centerline{\setlength{\unitlength}{1pt}
\centering
\begin{picture}(255,170)(0,0)
\put(2,0){
\includegraphics[width=8.2cm]{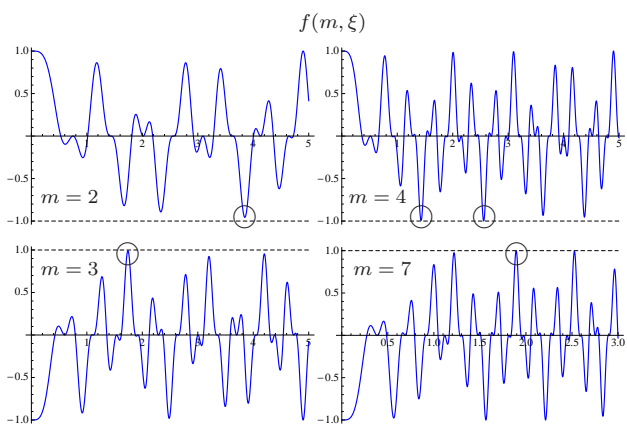}}
\end{picture}}
\caption{Plots of $f\!\left(m,\xi\right)$ against $\xi$ for $m=2,3,4,\;\mathrm{and}\;7$. The circled points denote locations where good approximations of a CZ gate can be obtained; these are detailed in table~\ref{table:1}.}\label{fig:fnplots}
\end{figure}

In the regime where the coupling is not far-off-resonant (\textit{i.e.},
$|\delta/\Omega|\leq1$), the populations of the states oscillate. To prevent leakage of the total population from the subspace spanned by the four ground states,  we need precise control on the applied pulse duration $t$ such that
$\sin\left(\Omega^{(1)} t/2\right)=\sin\left(\Omega^{(2)} t/2\right)=0$,
\textit{i.e.}, no population is left in the Rydberg states. Under this condition, the gate is
\begin{equation}\label{eq:GeneralGate}
G(t)\widehat{=}{\left(\begin{array}{@{}cccc@{}} 
e^{\frac{i}{2}\delta t}&0&0&0\\
0&\mathbox{4em}{c}{\cos{\left(\frac{1}{2}\Omega^{(1)} t\right)}}&0&0\\
0&0&\mathbox{4em}{c}{\cos{\left(\frac{1}{2}\Omega^{(1)} t\right)}}&0\\
0&0&0&\mathbox{5.5em}{c}{\cos{\left(\frac{1}{2}\Omega^{(2)} t\right)}}
\end{array}\right)}.
\end{equation}
We restrict the gate operation time to
\begin{eqnarray}
T_\mathrm{G}=2m\pi/|\delta|
\end{eqnarray}
for any positive integer $m$.
This allows two different realizations of a CZ gate:
$C_Z=\mathrm{diag}\{1,1,1,-1\}$, where the sign flip is on $\ket{11}$, which requires
\begin{eqnarray}
\cos(m\pi \Omega^{(1)}/\delta)=-\cos(m\pi\Omega^{(2)}/\delta)=(-1)^{m}\,;
\end{eqnarray}
or $C_Z=\mathrm{diag}\{-1,1,1,1\}$, where the sign flip is on $\ket{00}$, which requires
\begin{eqnarray}
\cos(m\pi \Omega^{(1)}/\delta)=\cos(m\pi\Omega^{(2)}/\delta)=(-1)^{m+1}\,.
\end{eqnarray}
We note that instead of coupling $\ket{1}$ to $\ket{r}$ while leaving $\ket{0}$ decoupled, one can couple $\ket{0}$ to a Rydberg state $\ket{r'}$ while leaving $\ket{1}$ decoupled. By doing so, the conditions for the two CZ gates are swapped. Thereby, fulfilling either one of the conditions, both $C_Z$ operations can be realized (the labelling of the computational basis states $\ket{0}$ and $\ket{1}$ is anyway, very often, interchangeable). Hence, the hunt for solutions follows one simple rule,
\begin{equation}\label{eq:f}
f(m,\xi) \equiv {\left[\cos\big(m\pi\Omega^{(1)}/\delta\big)\right]}^2\cos\big(m\pi\Omega^{(2)}/\delta\big)=(-1)^{m+1},
\end{equation}
where $\xi\equiv|\Omega|/|\delta|$. There are no exact solution to this equation, but one can find approximate solutions that are good enough for practical use.

We search for approximate solutions of eq.~(\ref{eq:f}) by first fixing the integer $m$, and then for each fixed $m$, finding values of $\xi$ such that $f\!\left(m,\xi\right)$ is very close to $(-1)^{m+1}$; see fig.~\ref{fig:fnplots}.
Five of the good approximate solutions are listed in table~\ref{table:1}.
For $m\!=\!3$, a gate fidelity $F_\mathrm{min}>0.99$ can be achieved with a pulse duration of about six Rabi-oscillation cycles of the $\ket{01}\leftrightarrow\ket{0r}$ transition. Without extending the pulse duration by much, higher gate fidelities can be achieved for $m\!=\!4$. The last row of the table lists an excellent solution for $m\!=\!7$ that offers a gate fidelity of 0.9990 in the ideal case. Solutions with even higher gate fidelities can be found for larger $m$ and/or $\xi$. However, because of their correspondingly longer pulse durations, it might not be practical to use some of these solutions that offer marginally higher fidelity. In practice, the choice of solution will depend on the specific experimental set-up and the type of noise encountered.
\begin{table}[t]
{\renewcommand{\arraystretch}{1.1}
\renewcommand{\tabcolsep}{3.5pt}
\caption{A list of five good approximate solutions for the CZ gate.  $\tau_{j}=2\pi/|\Omega^{(j)}|$ are the Rabi-oscillation periods for the excitation of one or two atoms. $F_\mathrm{min}$ and $\widetilde F_\mathrm{min}$ are the minimum gate fidelities over all initial two-qubit states with $\Delta_{rr}\rightarrow\infty$ and ${\Delta_{rr}=8}$\,GHz, respectively. $\widetilde F_{\textsc{D},\mathrm{min}}$ is the estimated minimum fidelity with $\Delta_{rr}=8$\,GHz and a Doppler shift of standard deviation ${\Delta_{\textsc{D}}=100}$\,kHz for each of the two atoms~\cite{DopplerError}. The fourth digit in $\xi$ corresponds to a variation of $\sim\!\!10$\,kHz in $\delta$. In an experiment with fluctuating values of $\Omega$ and $\delta$, the average values of $\Omega$ and $\delta$ should be adjusted according to $\xi$.}\label{table:1}
\begin{tabular}{@{}cccccccc@{}} \hline\hline\rule[-1.6ex]{0pt}{18pt}%
$m$&$\xi$&$f\left(m,\xi\right)$&$T_\mathrm{G}/\textcolor{black}{\tau_1}$&$T_\mathrm{G}/\textcolor{black}{\tau_2}$&$F_\mathrm{min}$&$\widetilde{F}_\mathrm{min}$&$\widetilde F_{\textsc{D},\mathrm{min}}$\\ [-1pt]
\hline
2 & 3.840 & -0.9707 &  7.94 & 11.00 &0.9633 &0.9633 & 0.9598\\ 
3 & 1.743 & \hphantom{-}0.9941 & 6.03 & 7.98 & 0.9938 &0.9938 & 0.9920\\ 
4 & 1.428 & -0.9955 & 6.98 & 9.02 &0.9948 &0.9948 & 0.9921\\ 
4 & 2.558 & -0.9983 & 10.99 & 15.01 &0.9979 &0.9969 & 0.9898\\ 
7 & 1.894 & \hphantom{-}0.9985 & 14.99 & 20.01 &0.9990 &0.9973 & 0.9853\\ 
\hline\hline
\end{tabular}}
\end{table}

\section{An example}
\begin{figure}[t]
\centerline{\setlength{\unitlength}{1pt}
\centering
\begin{picture}(220,272)(0,0)
\put(-2,-2){\includegraphics[width=7.6cm]{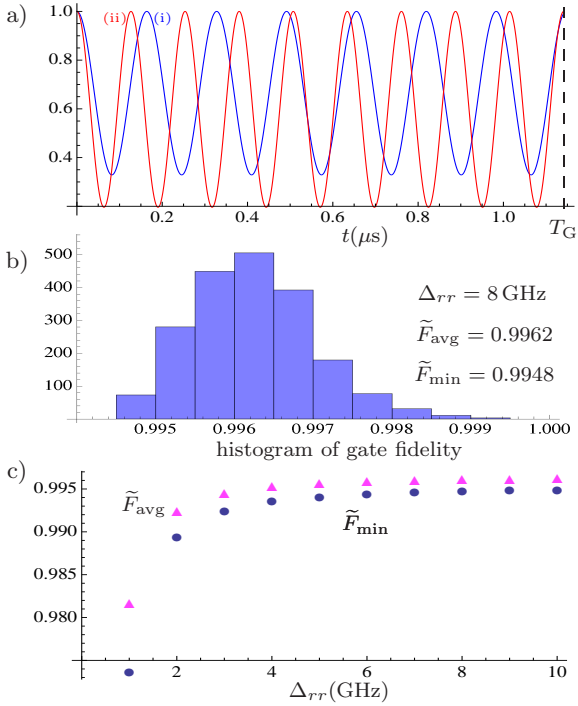}}
\end{picture}}
\caption{a) Population of states using the third solution listed in table~\ref{table:1} with $\Omega\!=\!5(2\pi)\,$MHz, $\delta\!=\!3.5(2\pi)\,$MHz and $\Delta_{rr}\!=\!8\,$GHz. The blue (i) and red (ii) curves show the population of $\ket{1}$ and $\ket{11}$ under Hamiltonians $H_I^{(1)}$ and $H_I^{(2)}$ for a pulse duration of
$T_\mathrm{G}=1.143$\,$\mu$s. b) For the same parameters as in a), the histogram of the gate fidelity for 2000 randomly generated initial states. c) A plot of the minimum fidelity $\widetilde F_\mathrm{min}$ and the average fidelity $\widetilde F_\mathrm{avg}$ against the Rydberg blockade energy $\Delta_{rr}$.}\label{fig:fig3}
\end{figure}

We now take a closer look at the third solution in table~\ref{table:1} with $m=4$ and $\xi=1.428$. Assuming coupling strength $\Omega=5(2\pi)\,$MHz and Rydberg blockade energy $\Delta_{rr}=8\,$GHz, the required detuning is $\delta=\Omega/1.428=3.50(2\pi)\,$MHz and the gate operation time is $T_\mathrm{G}=1.143\,\mu$s. Because the transition is off-resonant, the amplitude of the Rabi oscillation between $\ket{1}$ and $\ket{r}$ is less than 1; see fig.~\ref{fig:fig3}a. The simulated gate operator is
\begin{equation}
G(T_\mathrm{G})\widehat=\mbox{\small$-\!\!\left(\!\!\begin{array}{cccc} -1&0&0&0\\0&\!\!0.997+0.045i\!\!\!&0&0\\0&0&\!\!\!0.997+0.045i\!\!\!\!&0\\0&0&0&\!\!\!0.998+0.051i\!\!\!\end{array}\right)$}.\label{eq:10}
\end{equation}
This gate is not unitary mainly because, for all approximate solutions, the light pulse does not stop exactly at full Rabi cycles and a tiny fraction of population is left in the Rydberg states, contributing to the imperfection of the gate. 
By choosing a suitable Rydberg state, the blockade energy $\Delta_{rr}$ can be a few GHz, which is much larger than the atom-field coupling $\Omega$, so that the net effect of the imperfect Rydberg blockade is negligible.

The fidelity histogram of the gate $G(T_\mathrm{G})$ in eq.~(\ref{eq:10}) for a sample of 2000 randomly generated initial states (pure states with numerically generated random complex population amplitudes from uniform distributions) is shown in fig.~\ref{fig:fig3}b. The average gate fidelity is ${\widetilde F_\mathrm{avg}=0.9962}$ and the minimum fidelity is ${\widetilde F_\mathrm{min}=0.9948}$.

\section{Extension to an arbitrary Cphase gate}

With the general expression in {eq.~(\ref{eq:GeneralGate})}, the scheme can be adapted to an arbitrary Cphase gate $C_\mathrm{ph}(\phi)\!=\!\mathrm{diag}\{e^{i\phi},1,1,1\}$. To do this, we set the gate operation time $T_\mathrm{G}\!=\!2(m\pi+\phi)/|\delta|$, where $m$ is a positive integer. The gate is obtained if $g(m,\xi)\!\equiv\!\cos[(m\pi+\phi)\Omega^{(1)}/\delta]+\cos[(m\pi+\phi)\Omega^{(2)}/\delta]=2(-1)^m$. 
The construction of such gates is similar to that of a CZ gate (a special case of a Cphase gate), thus we omit the details and just illustrate it by an example. 
For $\phi\!=\!2\pi/3$, one approximate solution is $m\!=\!2$ and $\xi\!=\!2.00$. With the parameters for the example in fig.~\ref{fig:fig3} (\textit{i.e.}, $\Omega\!=\!5(2\pi)\,$MHz and $\Delta_{rr}\!=\!8\,$GHz), we need the detuning $\delta\!=\!2.5(2\pi)\,$MHz and the gate operation time $T_\mathrm{G}\!=\!1.069\,\mu$s. In this case, the average gate fidelity is $\widetilde F_\mathrm{avg}=0.9960$ and the minimum fidelity is $\widetilde F_\mathrm{min}=0.9944$. Better approximate solutions can be found with larger $m$ values. Following the same procedure, such high gate fidelity can be obtained for any value of $\phi$.

\section{Robustness}

The imperfections of practical quantum gates arise from errors of two kinds: intrinsic errors and technical errors. The average and minimum gate fidelities given in the previous sections take the intrinsic errors into account. These intrinsic errors are due to the choice of approximate solutions and finite Rydberg blockade. The former error ($\sim\!\! 10^{-3}$ in the given examples) can be further reduced by choosing better solutions of the equations; in practice, this might not be worth the trouble. The latter error can be below $10^{-4}$ for Rydberg levels with principal quantum number $n\!>\!100$ and a large Rydberg blockade energy~\cite{Saffman11, Walker08}. Table~\ref{table:1} shows that the effect of finite Rydberg blockade is slightly stronger for the solutions with higher $m$ values. 
One only requires $|\Delta_{rr}|\gg|\Omega|$, \textit{i.e.}, $|\Delta_{rr}|$ should be at least about a hundred times larger than $\Omega$ for it to be completely negligible; see fig.~\ref{fig:fig3}c.
For the solution with $\Omega\!=\!5(2\pi)\,$MHz and $\delta\!=\!3.5(2\pi)\,$MHz, a Rydberg blockade energy $\Delta_{rr}$ greater than 2\,GHz is required to achieve a minimum gate fidelity of higher than 0.99. With small interatomic distance, a Rydberg blockade energy of this order is experimentally achievable~\cite{PhysRevA.75.032712, Walker08, PhysRevLett.112.183002}, \textcolor{black}{and, if necessary, the atoms can be pulled apart to implement single-qubit gates~\cite{Jau16}.}

\begin{figure}[t]
\centerline{\setlength{\unitlength}{1pt}
\centering
\begin{picture}(230,340)(0,0)
\put(0,0){\includegraphics[width=7.8cm]{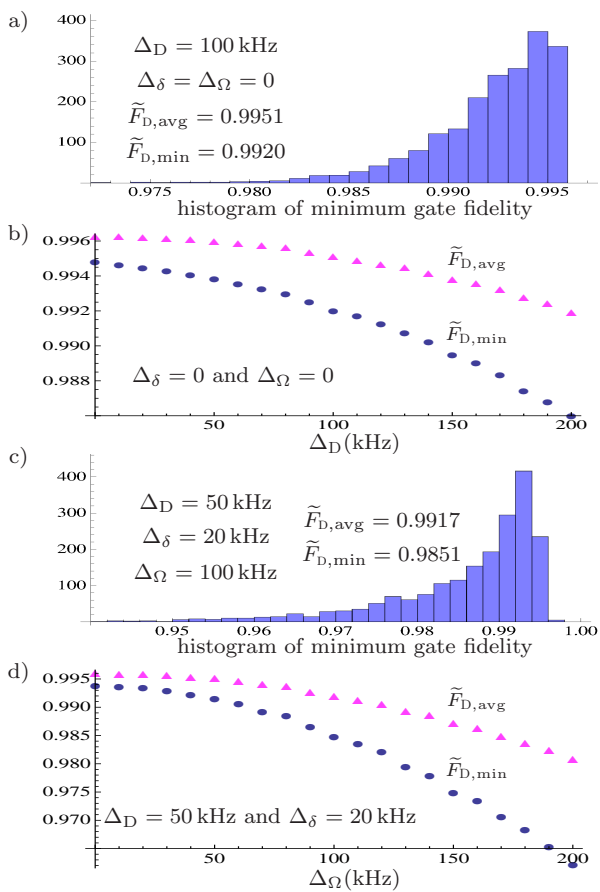}}
\end{picture}}
\caption{Robustness of the CZ gate for the third solution listed in table~\ref{table:1} with $\Omega\!=\!5(2\pi)\,$MHz, $\delta\!=\!3.5(2\pi)\,$MHz, $\Delta_{rr}\!=\!8\,$GHz and $T_\mathrm{G}=1.143\,\mu$s. a) The histogram of the estimated minimum gate fidelity $\widetilde F_{\textsc{D},\mathrm{min}}$ for 2000 simulations. For each simulation, the Doppler shift is randomly generated from a Gaussian distribution with standard deviation $\Delta_{\textsc{D}}=100\,$kHz; the estimated minimum gate fidelity is the average of 2000 values obtained for the individual simulations. b) The estimated minimum gate fidelity $\widetilde F_{\textsc{D},\mathrm{min}}$ and the average gate fidelity $\widetilde F_{\textsc{D},\mathrm{avg}}$ (averaged over random initial states) against the standard deviation $\Delta_{\textsc{D}}$ of the Doppler shift. For each $\Delta_{\textsc{D}}$, the values are averaged over 2000 simulations. c) The histogram of the estimated minimum gate fidelity $\widetilde F_{\textsc{D},\mathrm{min}}$ for 2000 simulations with stochastic noise in $\delta$, $\Omega$ and Doppler shift $\Delta_{\textsc{D}}$. The distributions of the noise are Gaussian with standard deviation $\Delta_\delta=20$\,kHz, $\Delta_\Omega=100$\,kHz and $\Delta_{\textsc{D}}=50\,$kHz, respectively. d) The simulated minimum and average gate fidelities against $\Delta_\Omega$.}\label{fig:fig4}
\end{figure}

Technical errors arise from spontaneous emission during the Rydberg excitation, Doppler broadening, and other experimental imperfections. For a Rydberg $\pi$ pulse, the error due to spontaneous emission is of the order of $10^{-4}$~\cite{Saffman10}. Although in our scheme the atoms undergo a few Rabi cycles during the gate operation time, the probability of Rydberg excitation is largely suppressed by the off-resonant light field ($|\Omega/\delta|\!\sim\!O(1)$); see fig.~\ref{fig:fig3}a. Thus, the error due to spontaneous emission should be no more than $10^{-4}$. Doppler broadening affects the detuning $\delta$, and thus the gate fidelity. For a $^{87}$Rb atom at 75\,$\mu$K, the energy shift due to the Doppler effect is about 40\,kHz, if the Rydberg excitation is achieved by two counter-propagating Raman beams via the 5$p$ state. As shown by fig.~\ref{fig:fig4}a, our scheme is not very sensitive to Doppler shift errors because of the large detuning $|\delta|\gg\Delta_{\textsc{D}}$. An average gate fidelity higher than 0.99 can be achieved at 75\,$\mu$K for a stochastic Doppler shift with standard deviation $\Delta_{\textsc{D}}=100$\,kHz, while the gate fidelity for the adiabatic scheme of Ref.~\cite{Jaksch00} is below 96\% because of the Doppler effect~\cite{Muller14}. Even for a Doppler-free configuration, the decoherence of the adiabatic scheme is dominated by the interatomic dipole forces of an imperfect blockade and population loss owing to the non-adiabatic evolution~\cite{Keating15}, which are not pertinent issues for our scheme. 

Other experimental imperfections may also affect the accuracies of $\delta$, $\Omega$ and the gate duration $T_\mathrm{G}$. 
Although $\delta$ can be controlled with an accuracy of 10\,kHz or even smaller, controlling the Rabi frequency at this level is not easy. The Rabi frequency $\Omega$ depends on the spatial uniformity and stability of the lasers as well as the distance between the two atoms, which are, in practice, more challenging to control precisely than the laser detuning. Yet, our scheme is not so sensitive to errors in $\delta$ and $\Omega$:
Numerical simulations show that an average fidelity of 0.99 can be maintained when $\delta$, $\Omega$ and the Doppler shift $\Delta_{\textsc{D}}$ have stochastic deviations drawn from Gaussian distributions with standard deviations of 20\,kHz, 100\,kHz and 50\,kHz, respectively; see fig.~\ref{fig:fig4}b.

\begin{figure}[t]
\centerline{\setlength{\unitlength}{1pt}
\centering
\begin{picture}(235,140)(0,0)
\put(2,-2){\includegraphics[width=7.7cm]{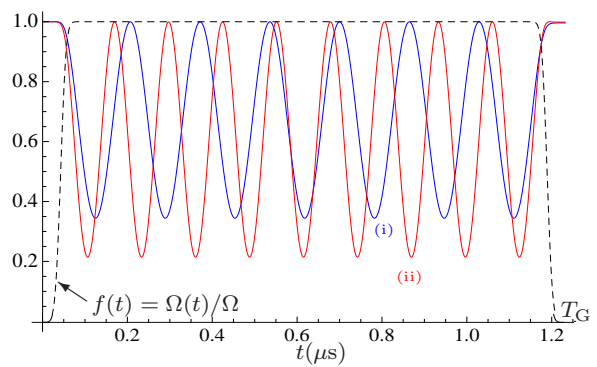}}
\end{picture}}
\caption{Population of states with time-dependent switching-one-and-off Rabi coupling. The blue (i) and red (ii) curves show the population of $\ket{1}$ and $\ket{11}$ under Hamiltonians $H_I^{(1)}$ and $H_I^{(2)}$. The dashed black curve shows the time-dependent function of the Rabi coupling strength: $f(t)=\Omega(t)/\Omega$, where $\Omega(t)$ is given in eq.~(\ref{eq:Omegat}). Parameters used for this simulation are $\Omega=5(2\pi)$\,MHz, $\xi=1.430$, $\Delta_{rr}=8$\,GHz, $\Delta_T=10$\,ns and $T_\mathrm{G}=1.228\,\mu$s, and a gate fidelity $\widetilde F_\mathrm{min}=0.994$ is obtained.}\label{fig:fig5}
\end{figure}

In this analysis, the gate time $T_G$ is estimated using square pulses with instantaneous switching-on-and-off. However, experimental laser pulses have time-dependent switching profiles depending on the AOM/EOM apparatus. Since the actual pulse shape is the same from realization to realization with very high precision, the parameters can be optimized for the time-dependent pulse profile to achieve high gate fidelities. 
Figure~\ref{fig:fig5} shows an example where the time-dependent atom-light coupling strength is modelled by 
\begin{eqnarray}
\Omega(t)=\left\{\!\!\begin{array}{l}\frac{1}{2}\Omega\!\left[1+\mathrm{erf}\left(\frac{t}{\sqrt{2}\Delta_T}-3\right)\right]\;\;\mathrm{for}\;\;t<\frac{1}{2}T_\mathrm{G}\,,\\[2ex] \frac{1}{2}\Omega\!\left[1+\mathrm{erf}\left(\frac{T_\mathrm{G}-t}{\sqrt{2}\Delta_T}-3\right)\right]\;\;\mathrm{for}\;\;t\geq\frac{1}{2}T_\mathrm{G}\,.\end{array}\right.\label{eq:Omegat}
\end{eqnarray}
For $\Delta_T=10$\,ns --- giving rise to a switching time of about 60\,ns --- a gate fidelity $\widetilde F_\mathrm{min}=0.994$ can be obtained with $\Omega=5(2\pi)$\,MHz, $\xi=1.430$, $\Delta_{rr}=8$\,GHz and $T_\mathrm{G}=1.228\,\mu$s.
Such an optimization can also be done for other types of time-dependent pulses.

All existing schemes, including the scheme presented in this letter, require the two atoms to be uniformly coupled to the addressing lasers. The uniformity of the Rabi frequencies for the two atoms depends on the atomic distance and the laser configuration. Since our scheme does not require individual addressing of the atoms, we do not have any constraint on the atomic distance --- other than that the atoms remain within the Rydberg blockade radius --- and a spatially uniform Rabi coupling can be achieved when the atoms are close to each other. The laser configuration depends on the experimental choice of the Rydberg coupling scheme. Rydberg excitation can be achieved by a direct coupling where Doppler shifts can be suppressed when using two lasers~\cite{Keating15}, by a two-photon Raman transition where the Doppler error is reduced with counter-propagating beams, or by a three-photon transition where a Doppler-free configuration can be obtained by adjusting the angles between the laser beams~\cite{PhysRevA.84.053409}. Different laser configurations have different sensitivity to beam pointing and intensity fluctuations, and each configuration must be assessed on its own. Moreover, the presence of the intermediate states for multi-photon transitions would affect slightly the phases accumulated during the transition, especially when the detuning to the intermediate states is not large enough.
We shall report a more detailed error analysis in a technical article. 

\section{Conclusion}
Our novel scheme for neutral-atom Cphase gates employs only a single Rydberg pulse, which addresses the two atoms simultaneously and symmetrically and operates for a duration of a few Rabi-oscillation cycles. All other schemes are designed to give perfect gates under ideal circumstances and, as a consequence, are complicated and suffer much when the circumstances are non-ideal. In marked contrast, we accept right away that any implementation will have imperfections and design a slightly imperfect but much more robust scheme which is also relatively easy to implement. An analysis of both intrinsic errors and technical errors shows that Cphase gates with fidelities higher than 0.99 can be constructed with current technology.

\acknowledgments
We acknowledge helpful discussions with Wenhui Li, Luyao Ye and Nana Siddharth. This work is funded by the Singapore Ministry of Education and the National Research Foundation of Singapore. H. K. N is also funded by a Yale-NUS College start-up grant.


\clearpage

\end{document}